\documentclass[twocolumn,tighten,times]{aastex631}
\accepted{\today}

\hypersetup{linkcolor=blue,citecolor=blue,filecolor=blue,urlcolor=magenta}
\newcommand{\um}{$\mu$m}
\definecolor{navyblue}{RGB}{0,50,250}

\newcommand{\co}{CO}
\newcommand{\lineco}{$^{12}$\co(2--1)}
\shorttitle{The magnetic fields in the starburst ring of NGC1097}
\shortauthors{Lopez-Rodriguez et al.}
\begin{document}

\title{Extragalactic magnetism with SOFIA (Legacy Program) - II: \\
A magnetically-driven flow in the starburst ring of NGC~1097\footnote{The SOFIA Legacy Program for Magnetic Fields in Galaxies provides a software repository at \url{https://github.com/galmagfields/hawc}, and publicly available data at \url{http://galmagfields.com/}}}

\correspondingauthor{Lopez-Rodriguez, E.}
\email{elopezrodriguez@stanford.edu}

\author{Enrique Lopez-Rodriguez}
\affil{Kavli Institute for Particle Astrophysics \& Cosmology (KIPAC), Stanford University, Stanford, CA 94305, USA}

\author{Rainer Beck}
\affil{Max-Planck-Institut f\"ur Radioastronomie, Auf dem H\"ugel 69, 53121 Bonn, Germany}

\author{Susan E. Clark}
\affil{Kavli Institute for Particle Astrophysics \& Cosmology (KIPAC), Stanford University, Stanford, CA 94305, USA}

\author{Annie Hughes}
\affil{Institute for Research in Astrophysics \& Planetology, CNRS, France}

\author{Alejandro S. Borlaff}
\affil{NASA Ames Research Center, Moffett Field, CA 94035, USA}

\author{Evangelia Ntormousi}
\affil{Scuola Normale Superiore, Piazza dei Cavalieri 7, 56126 Pisa, Italy}

\author{Lucas Grosset}
\affil{Kavli Institute for Particle Astrophysics \& Cosmology (KIPAC), Stanford University, Stanford, CA 94305, USA}

\author{Konstantinos Tassis}
\affil{Institute of Astrophysics, Foundation for Research and Technology-Hellas, Vasilika Vouton, 70013 Heraklion, Greece}
\affil{Department of Physics, and Institute for Theoretical and Computational Physics, University of Crete, 70013 Heraklion, Greece}

\author{John E. Beckman}
\affil{Instituto de Astrof\'isica de Canarias, C/ Via L\'actea s/n, 38200 La Laguna, Tenerife, Spain}
\affil{Departamento de Astrof\'isica, Universidad de La Laguna, Avda. Astrofísico Fco. S\'anchez s/n, 38200 La Laguna, Tenerife, Spain}

\author{Kandaswamy Subramanian}
\affil{Inter-University Centre for Astronomy and Astrophysics, Post Bag 4, Ganeshkhind, Pune 411007, India}

\author{Daniel Dale}
\affil{Department of Physics \& Astronomy, University of Wyoming, Laramie, WY, USA}

\author{Tanio D\'iaz-Santos}
\affil{Institute of Astrophysics, Foundation for Research and Technology-Hellas (FORTH), Heraklion, 70013, Greece}

%%%%%%%%%%%%%%%%%%%%
%%%%% ABSTRACT %%%%%
%%%%%%%%%%%%%%%%%%%%

\begin{abstract} 
Galactic bars are frequent in disk galaxies and they may support the transfer of matter towards the central engine of active nuclei. The barred galaxy NGC~1097 has magnetic forces controlling the gas flow at several kpc scales, which suggest that magnetic fields (B-fields) are dynamically important along the bar and nuclear ring. However, the effect of the B-field on the gas flows in the central kpc scale has not been characterized. Using thermal polarized emission at 89 \um~with HAWC+/SOFIA, here, we measure that the polarized flux is spatially located at the contact regions of the outer-bar with the starburst ring. The linear polarization decomposition analysis shows that the $89$ \um\ and radio ($3.5$ and $6.2$ cm) polarization traces two different modes, $m$, of the B-field: a constant B-field orientation and dominated by $m=0$ at $89$~\um, and a spiral B-field dominated by $m=2$ at radio. We show that the B-field at 89 \um~is concentrated in the warmest region of a shock driven by the galactic-bar dynamics in the contact regions between the outer-bar with the starburst ring. Radio polarization traces a superposition of the spiral B-field outside and within the starburst ring. According to Faraday rotation measures between $3.5$ and $6.2$ cm, the radial component of the B-field along the contact regions points toward the galaxy’s center on both sides. We conclude that gas streams outside and within the starburst ring follow the B-field, which feeds the black hole with matter from the host galaxy.
\end{abstract}

%%%%%%%%%%%%%%%%%%%%%%%%
%%%%% INTRODUCTION %%%%%
%%%%%%%%%%%%%%%%%%%%%%%%

\section{Introduction} \label{sec:INT}

Kinematic studies using several gas tracers (e.g. CO, HCN) of the interstellar medium (ISM) in galaxies have shown streaming motions from the bar towards the active nuclei associated with spiral structures. These results are commonly interpreted as gas inflows fueling the active nuclei from kpc scales through the starburst ring \citep{Kohno2003,Fathi2006,Prieto2019}. In a hydrodynamical (HD) framework, the gas flow along the galactic bar suffers a large deflection angle driven by the gravitational potential of the bar. As gas is dissipative, it is shocked and looses angular momentum, which creates a radial component. The gas flow then transition from the gas lane to a new orbit, producing a ring and/or central spiral structures around the nucleus of the galaxy \citep{Athanassoula1992,Athanassoula1992b,Piner1995}. The gas can subsequently collapse, increase the density within the ring, and form a starburst ring \citep[i.e.][]{Athanassoula1992,Athanassoula1992b,Sormani2015}. High density regions that correspond with sharp changes in gas velocities and temperature are typically identified as a shock driven by the galactic bar. The gas is shocked at the orbits crossing from the galactic bar to the ring. Hereafter, we refer to these dense regions as `contact regions'. Note that this is a label considered in this manuscript to identify features of the same kinematic system produced by the bar potential. Although HD models can reproduce the kinematics of galactic bars and formation of starburst rings, these models have difficulty reproducing the gas inflows towards the nucleus of the galaxy. Thin bars with high axial ratios and without nuclear rings are required to reproduce the observed gas inflows \citep{Piner1995}. 

Magnetic fields (B-fields) have been found to be strong in the dust lanes and nuclear rings of barred galaxies \citep{Beck1999,Beck2002,Beck2005}, where magnetic forces can dominate the gas flows \citep{Beck1999,Beck2005}. These results suggest that the B-fields are dynamically important along the bar and nuclear rings of barred galaxies. Galactic bars generate shearing gas flows that stretch and amplify the B-field. A galactic bar provides a non-axisymmetric perturbation of the gravitational potential in a galaxy. \citet{CL1994} argued that regular B-fields may be enhanced by velocity gradients, and \citet{Moss1998} showed that dynamos can be affected by the presence of a galactic bar. Non-axisymmetric perturbations result in a B-field that rotates in a bar with high dynamo modes ($m=$ 1 or 2), where the resulting B-field may be a composition of ring-like and spiral structures towards the galaxy center \citep{Moss1998}. For an axisymmetric potential, the galactic dynamo predicts an azimuthal B-field mode of $m=$ 0 \citep[i.e.][]{CL1994,Moss1998}. In general, the non-axisymmetric gas flows in bars interact with B-fields and the magnetic stress removes angular momentum from the gas at the shocks. The dominant B-fields then deflect the gas flow from the galactic bar to a new orbit, producing a central ring and/or central spiral towards the nucleus. In this scenario, the deflection of the gas flow is driven by the transition between compressed B-field in the shocks to a magnetohydrodynamic (MHD) dynamo towards the nucleus. Two-dimensional MHD simulations \citep{Kim2012} have shown that for magnetized models of barred galaxies, features like shock waves at $\sim1$ kpc from the central black hole, MHD dynamos, and magnetic arms are indicative of the B-fields dominating the gas flows towards the central black hole. The gas flows probably follow the B-field, which feeds the black hole with matter from the host galaxy. MHD models can predict the observed gas inflow towards the central black hole with the combination of a bar and a ring. Thus, characterization of the observed B-field morphology (i.e. B-field modes) and direction at the location of these shocks (i.e. contact regions) provide the keys for understanding the gas flows towards the active nuclei from the galactic bar.

NGC~1097 \citep[$D=19.1$ Mpc, $1\arcsec = 92.6$ pc;][]{Willick1997} is typically classified as a barred spiral (SBb), which contains a low-luminosity active nucleus surrounded by a circumnuclear starburst ring of $\sim2$ kpc in diameter \citep{Hummel1987,GNC1988}. An inner bar at $\sim28^{\circ}$ is found within the starburst ring \citep{Quillen1995,Prieto2005}. \textit{Herschel} images show that the active nucleus does not contribute to the total far-infrared (FIR) emission in the central $2$ kpc \citep{Sandstrom2010}, in contrast with other nearby active galaxies \citep[i.e. Cygnus~A and NGC~1068;][]{LR2018a,ELR2018b}. The thermal emission from the starburst ring contributes up to 60\% of the total flux at 100 \um\ within the central $2$ kpc. The starburst ring is embedded in an outer bar of $\sim20$ kpc in diameter at an angle of $148^{\circ}$ and two spiral arms at larger scales \citep[see Fig.\,5 in][]{Quillen1995}. The dust lanes have low star formation rates and low opacity \citep{Quillen1995}.  

\begin{figure*}[ht!]
\includegraphics[angle=0,scale=0.43]{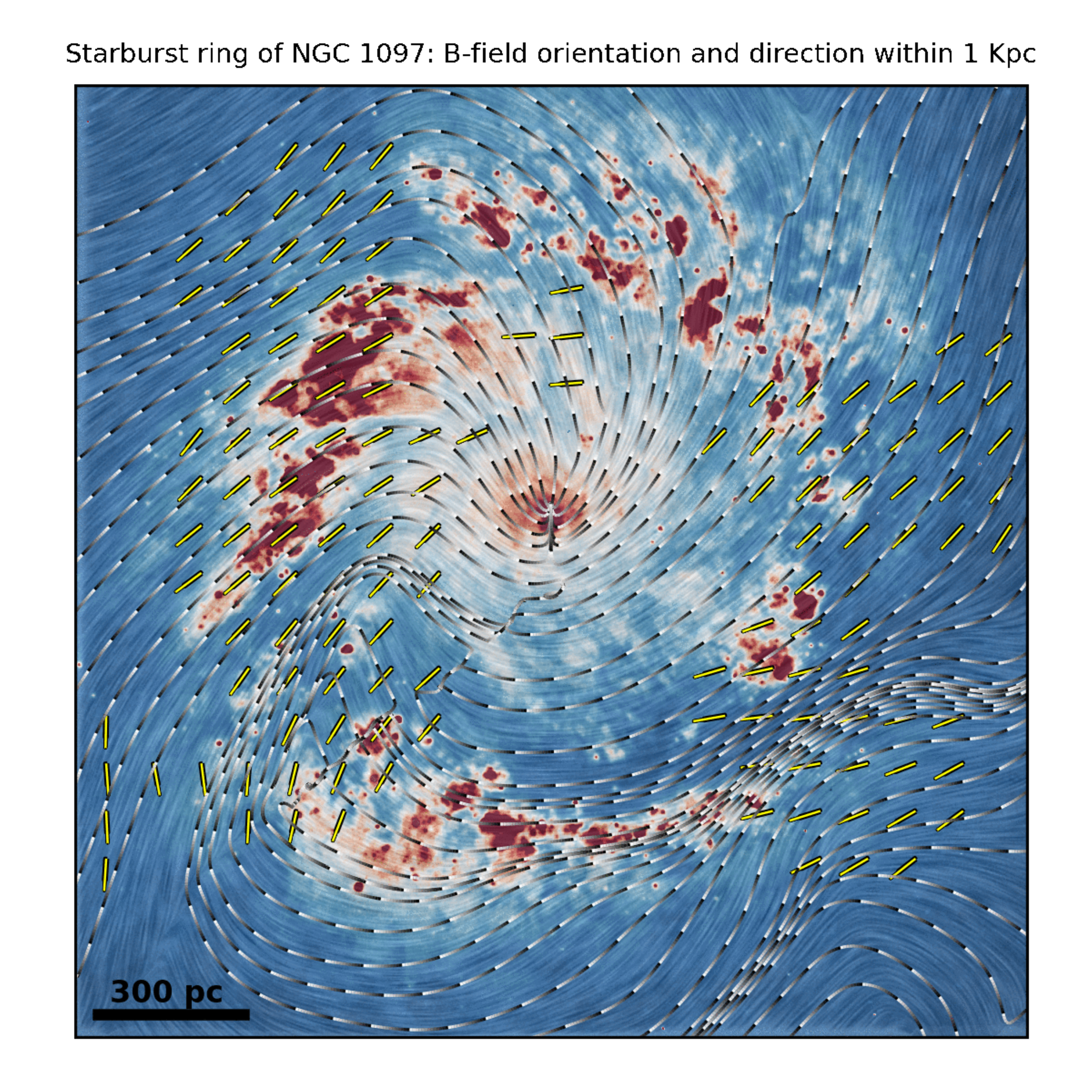}
\caption{B-field orientation and direction in the central 1 kpc starburst ring of NGC~1097. \textit{HST}/WFC3/F438W ultraviolet image (color scale), B-field orientation at $3.5$ cm (background streamlines), B-field direction at $3.5$ cm (black and white streamlines), and B-field orientation at $89$ \um\ (yellow lines) are shown. This figure illustrates the main results from this work. \label{fig:fig0}}
\epsscale{2.}
\end{figure*}

The equipartition B-field strength is estimated to be $\sim60~\mu$G in the starburst ring of NGC 1097 \citep{Beck1999,Beck2005}. The B-fields in the starburst ring spiral down towards the active nucleus at an angle of $\sim30^{\circ}$ (Fig. \ref{fig:fig0}), which is spatially coincident with the inner-bar at $\sim28^{\circ}$ \citep{Quillen1995,Prieto2005}. At larger scales, the gas streams follow the outer bar and then twist to follow the spiral arms at scales of several tens of kpc. The fact that the B-fields follow the spiral arms and the circumnuclear ring indicates the action of a large-scale galactic dynamo, which may be enhancing the B-field strength in this galaxy due to differential rotation. \citet{Beck1999} suggested that magnetic stress may be an efficient mechanism to fuel the central active nucleus in NGC1097. Further analysis of the thermal and non-thermal emission using radio polarimetric observations have shown that most of the star formation efficiency of the clouds in the starburst ring drops with increasing the B-field strength \citep{Tabatabaei2018}. The energy balance in the ISM of the staburst ring shows that the magnetic energy is in close equipartition with the turbulent kinetic energy. Both energies are a factor of ten higher than the thermal energy. These results imply that the starburst ring is magnetically critical, where the clouds are supported against the gravitational collapse. This results in inefficient high-mass star formation. Indeed, the starburst ring has a slightly lower star formation rate, $\sim2$ M$\odot$ yr $^{-1}$ \citep{Hsieh2011}, than the typical $3-11$ M$\odot$ yr $^{-1}$ in circumnuclear starbursts in barred galaxies \citep{Jogee2005}.

%%%%%%%%%%%%%%%%%
%%%% TABLE 1 %%%%
%%%%%%%%%%%%%%%%%
\begin{deluxetable*}{ccp{1.8cm}cp{1.8cm}ccccccc}[ht!]
\tablecaption{Summary of OTFMAP polarimetric observations. \emph{Columns, from left to right:} a) Observation date. b) Flight ID. c) Sea-level altitude during the observations (ft). d) Speed of the scan (\arcsec/sec). d) Phase of the Lissajous pattern ($^{\circ}$). f) Amplitudes in elevation (EL) and cross-elevation (XEL) of the scan (arcsec). g) Time per scan (s). h) Number of observation sets obtained (and rejected). i) On-source observation time (s).
\label{tab:table1} 
}
\tablecolumns{7}
\tablewidth{0pt}
\tablehead{\colhead{Date}	&	\colhead{Flight}	&	\colhead{Altitude}	&
\colhead{Scan Rate} & \colhead{Scan Phase} & \colhead{Scan Amplitude } & \colhead{Scan Duration} & \colhead{\#Sets (bad)} & \colhead{t$_{\mathrm{on-source}}$} \\ 
 \colhead{(YYYYMMDD)}	&		&	\colhead{(ft)}	&
\colhead{(\arcsec/sec)} & \colhead{($^{\circ}$)} & \colhead{ (EL $\times$ XEL; \arcsec)} & \colhead{(s)}  & & \colhead{(s)} \\
\colhead{(a)} & \colhead{(b)} & \colhead{(c)} & \colhead{(d)} & \colhead{(e)} & \colhead{(f)} & \colhead{(g)} & \colhead{(h)} & \colhead{(i)}
}
\startdata
	20200125	&	F653	&	38000-41000	&	100	&	-30, -23.3	&	90 $\times$ 90	& 120	&	3 	& 1440 \\
			&		&				&		& -10, -3.3, 3.3, 10.0, 17.0, 23.4, 30.0, -16.7	&	60$\times$60	&	120	&	6 (2)	&	2880	\\	
	20200128	&	F654	&	38000	&	100	&	-16.7	&	60$\times$60	&	120	&	5(1)	&	2400\\
\enddata
%\tablenotetext{a}{Measured and modeled photometry as described in Section \ref{subsec:pho}.}
\end{deluxetable*}
%%%%%%%%%%%%%%%%%%%

%%%%%%%%%%%%%%%%%%
%%%% FIGURE 0 %%%%
%%%%%%%%%%%%%%%%%%
%% The "ht!" tells LaTeX to put the figure "here" first, at the "top" next
%% and to override the normal way of calculating a float position
%%%%%%%%%%%%%%

%%%%%%%%%%%%%%%%%%
%%%% FIGURE 1 %%%%
%%%%%%%%%%%%%%%%%%
%% The "ht!" tells LaTeX to put the figure "here" first, at the "top" next
%% and to override the normal way of calculating a float position
\begin{figure*}[ht!]
\includegraphics[angle=0,scale=0.45]{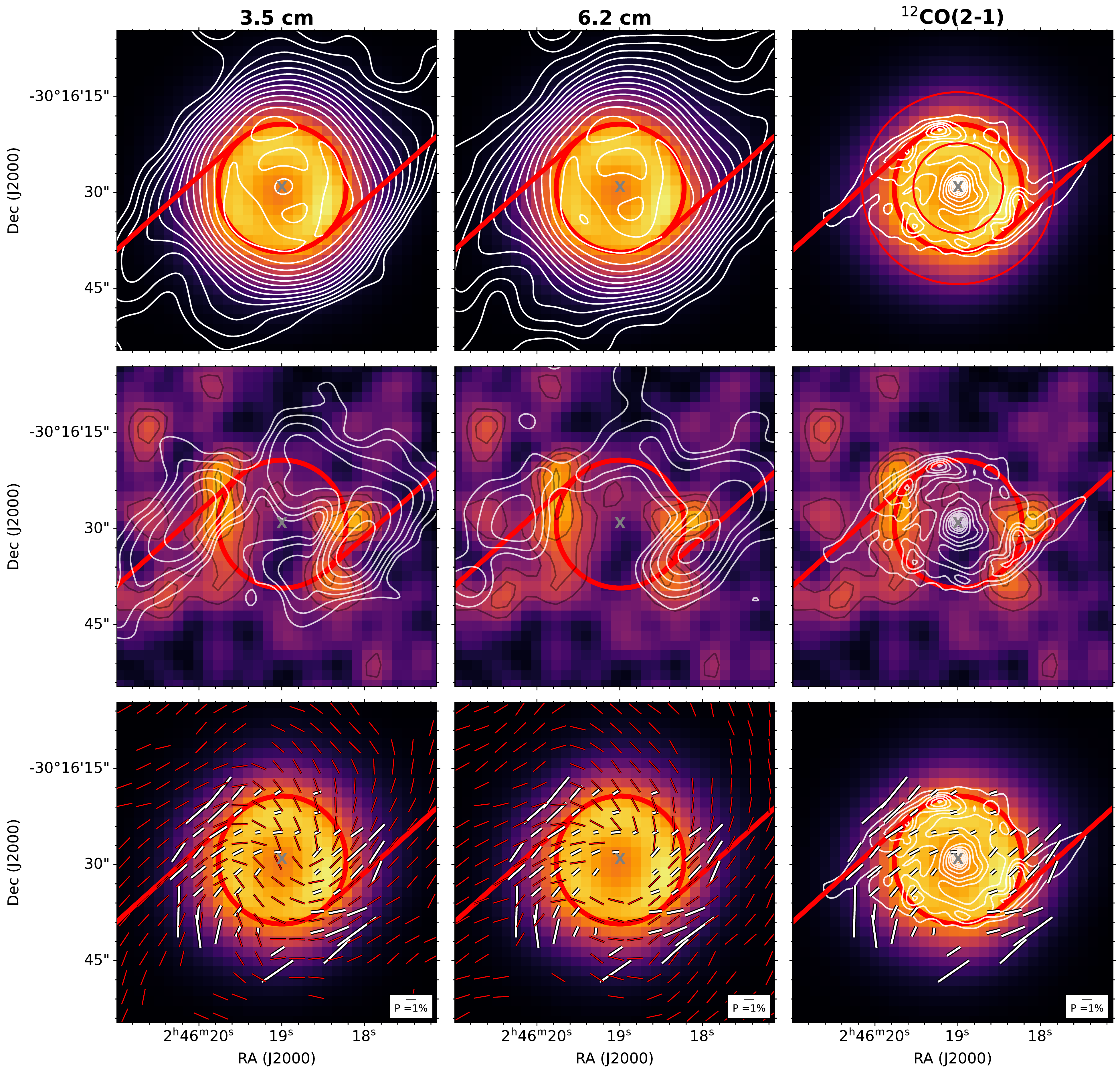}
\caption{Total and and polarized structures of NGC\,1097 as a function of wavelength. \textit{Top:} $89$ \um~total flux (colorscale) with overlaid $3.5$ cm (left) and $6.2$ cm (middle) total fluxes (white contours), and \lineco\ integrated emission line (right; white contours). 
\textit{Middle:} $89$ \um~polarized flux (colorscale) with overlaid $3.5$ cm (left) and $6.2$ cm (middle) polarized fluxes (white contours), and \lineco\ integrated emission line (right; white contours). 
\textit{Bottom:} $89$ \um~total flux (colorscale) with overlaid B-field orientation map at $89$ \um~(white lines) and radio wavelengths (red lines) at $3.5$ cm (left) and $6.2$ cm (middle). At $89$ \um, the length of the polarization shows the degree of polarization with a legend of $1$\% at the bottom right of the panel. For the polarization in radio, the lengths of the B-field orientations are normalized to unity.
In all figures, the orientations of the dust lane in the outer-bar (red solid lines) are parallel to the extended emission of the lowest contour of the \lineco\ integrated emission line. The central position of the starburst ring at a radius of $10\arcsec$ (red circle) and dust lanes (red solid lines) at a $PA\sim148^{\circ}$ are shown. For all panels, the AGN location (grey cross) is shown. The radial range used for the azimuthal profiles in Section \ref{subsec:RadAZ} is shown as red thin solid lines in the top-right panel.
\label{fig:fig1}}
\epsscale{2.}
\end{figure*}
%%%%%%%%%%%%%%

NGC~1097 offers one of the best laboratories for studying the B-field in the dense ISM of a barred galaxy.  Our goal is to characterize the morphology of the B-field inferred through magnetically aligned dust grains in the dense ISM of the central kpc of NGC~1097. By comparing the FIR polarimetric observations with the radio polarimetric observations and the kinematics of the molecular gas, we characterize the dust polarized emission and gas flows across the starburst ring. 

We describe in Section \ref{sec:OBS} the specifics of our observations. Section \ref{sec:ANA} shows the analysis of our observations with radio and molecular gas observations. The decomposition of the B-field morphology at 89 \um\ and radio observations, and the estimation of the B-field direction are shown in Section \ref{sec:FIRvsRadio}. Our discussions are described in Section \ref{sec:DIS} and our main conclusions are summarized in Section \ref{sec:CON}.

%%%%%%%%%%%%%%%%%%%%%%%%%%%%%%%%%%%%%%%%%%%
%%%%% OBSERVATIONS AND DATA REDUCTION %%%%%
%%%%%%%%%%%%%%%%%%%%%%%%%%%%%%%%%%%%%%%%%%%

\section{Observations and Data Reduction} \label{sec:OBS}

\subsection{HAWC+ observations}\label{subsec:HAWC_OBS}

NGC~1097 was observed (PI: Lopez-Rodriguez, E., ID: 07\_0034, and as part of the SOFIA Legacy Program\footnote{SOFIA Legacy Program: \url{http://galmagfields.com/}}) at $89$ \um~using the High-resolution Airborne Wideband Camera-plus \citep[HAWC+;][]{Vaillancourt2007,Dowell2010,Harper2018} on the $2.7$-m Stratospheric Observatory For Infrared Astronomy (SOFIA) telescope. HAWC+ polarimetric observations simultaneously measure two orthogonal components of linear polarization arranged in two arrays of $32 \times 40$ pixels each, with a pixel scale of  $4$\farcs$02$ pixel$^{-1}$, and beam size (full width at half maximum, FWHM) of $7$\farcs$80$ at $89$ \um.  We performed observations using the on-the-fly-map (OTFMAP) polarimetric mode. This technique is an experimental observing mode performed during SOFIA Cycle 7 observations as part of engineering time to optimize the polarimetric observations of HAWC+. This technique has been successfully applied to another galaxies, Centaurus A \citep{ELR2021} and Circinus (Grosset et al. in prep.), as well as the filamentary cloud L1495/B211 (Li et al. submitted to ApJ). Although we focus here on the scientific results of NGC~1097, what follows is the high-level steps of the OTFMAP polarimetric observations.

We performed OTFMAP polarimetric observations in a sequence of four Lissajous scans, where each scan has a different halfwave plate (HWP) position angle (PA) in the following sequence: $5^{\circ}$, $50^{\circ}$, $27.5^{\circ}$, and $72.5^{\circ}$. This sequence is called a `set' hereafter. In this new HAWC+ observing mode, the telescope is driven to follow a parametric curve with a nonrepeating period whose shape is characterized by the relative phases and frequency of the motion. Each scan is characterized by the scan amplitude, scan rate, scan angles, and scan duration.  A summary of the observations are shown in Table \ref{tab:table1}. 

We reduced the data using the Comprehensive Reduction Utility for SHARP II v.2.42-1 \citep[\textsc{crush};][]{kovacs2006,kovacs2008} and the \textsc{hawc\_drp\_v2.3.2} pipeline developed by the data reduction pipeline group at the SOFIA Science Center. Each scan was reduced using \textsc{crush}, which estimates and removes the correlated atmospheric and instrumental signals, solves for the relative detector gains, and determines the noise weighting of the time streams in an iterated pipeline scheme. Each reduced scan produces two images associated with each array. Both images are orthogonal components of linear polarization at a given HWP PA.  We estimated the Stokes IQU parameters using the double difference method in the same manner as the standard chop-nod observations carried by HAWC+ described in Section 3.2 by \citet{Harper2018}. The degree (P) and PA of polarization were corrected by instrumental polarization estimated using OTFMAP polarization observations of planets. We estimated a polarization uncertainty of $\sim0.8\%$. We estimated that the field-of-view (FOV) would rotate $\sim12^{\circ}$ across the first night of observations, which complicates the computation of the rotation of the Stokes parameters. To ensure the correction of the PA of polarization of the instrument with respect to the sky, we took each set with a fixed line-of-sight (LOS) of the telescope. These scan phase angles are shown in Table \ref{tab:table1}. For each set, we rotated the Stokes QU from the instrument to the sky coordinates. The polarization fraction was debiased and corrected by polarization efficiency. The final Stokes IQU, P, PA, polarized intensity (PI), and their associated errors were calculated and re-sampled to one-quarter of the beam size, $1\farcs95$ at $89$ \um.  Final images have a total elapsed time (overhead + on-source) of 7275\,s at $89$ \um, where 6720\,s corresponds to time on-source. For the OTFMAP polarization observations we estimate the overhead factor to be 1.08. Several sets, two the first night and one the second night, were removed due to tracking error issues during the observations.

%%%%%%%%%%%%%%%%%%
%%%% FIGURE 2 %%%%
%%%%%%%%%%%%%%%%%%
%% The "ht!" tells LaTeX to put the figure "here" first, at the "top" next
%% and to override the normal way of calculating a float position
\begin{figure*}[ht!]
\includegraphics[angle=0,scale=0.40]{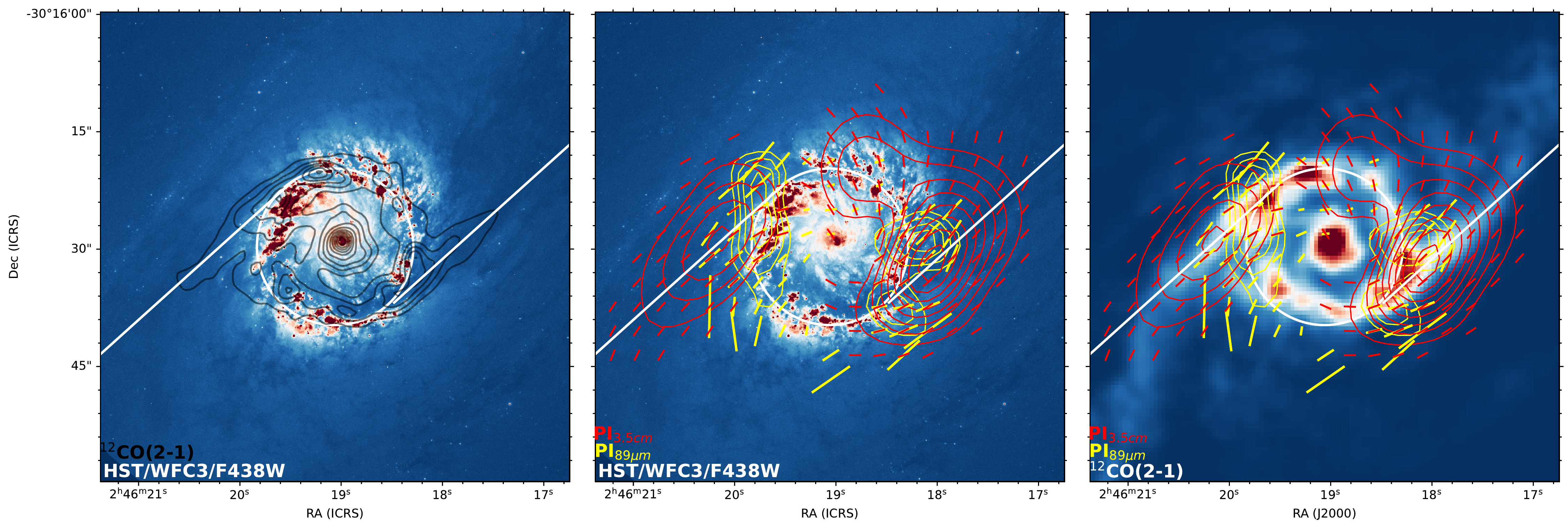}
\caption{B-fields in the diffuse (radio) and dense (FIR) ISM vs gas tracers and star-forming regions. Left: \textit{HST}/WFC3/F438W emission (color scale) and $^{12}$CO(2-1) integrated emission line (black contours).
Middle: \textit{HST}/WFC3/F438W emission (color scale) and polarized fluxes at 89 \um~(yellow contours) and 3.5 cm (red contours). The B-field orientation is shown using the color scheme at the bottom-left, where the B-field lines at 3.5 cm has been normalized to unity. 
Right: Same as the middle panel but with $^{12}$CO(2-1) integrated emission line (color scale). 
For all panels, the central position of the starburst ring at a radius of $10\arcsec$ (white circle) and dust lanes (white solid lines) at a PA$\sim148^{\circ}$ are shown.
\label{fig:fig2}}
\epsscale{2.}
\end{figure*}
%%%%%%%%%%%%%%%%%%

%%%%%%%%%%%%%%%%%%
%%%% FIGURE 3 %%%%
%%%%%%%%%%%%%%%%%%
%% The "ht!" tells LaTeX to put the figure "here" first, at the "top" next
%% and to override the normal way of calculating a float position
\begin{figure*}[ht!]
\includegraphics[angle=0,scale=0.59]{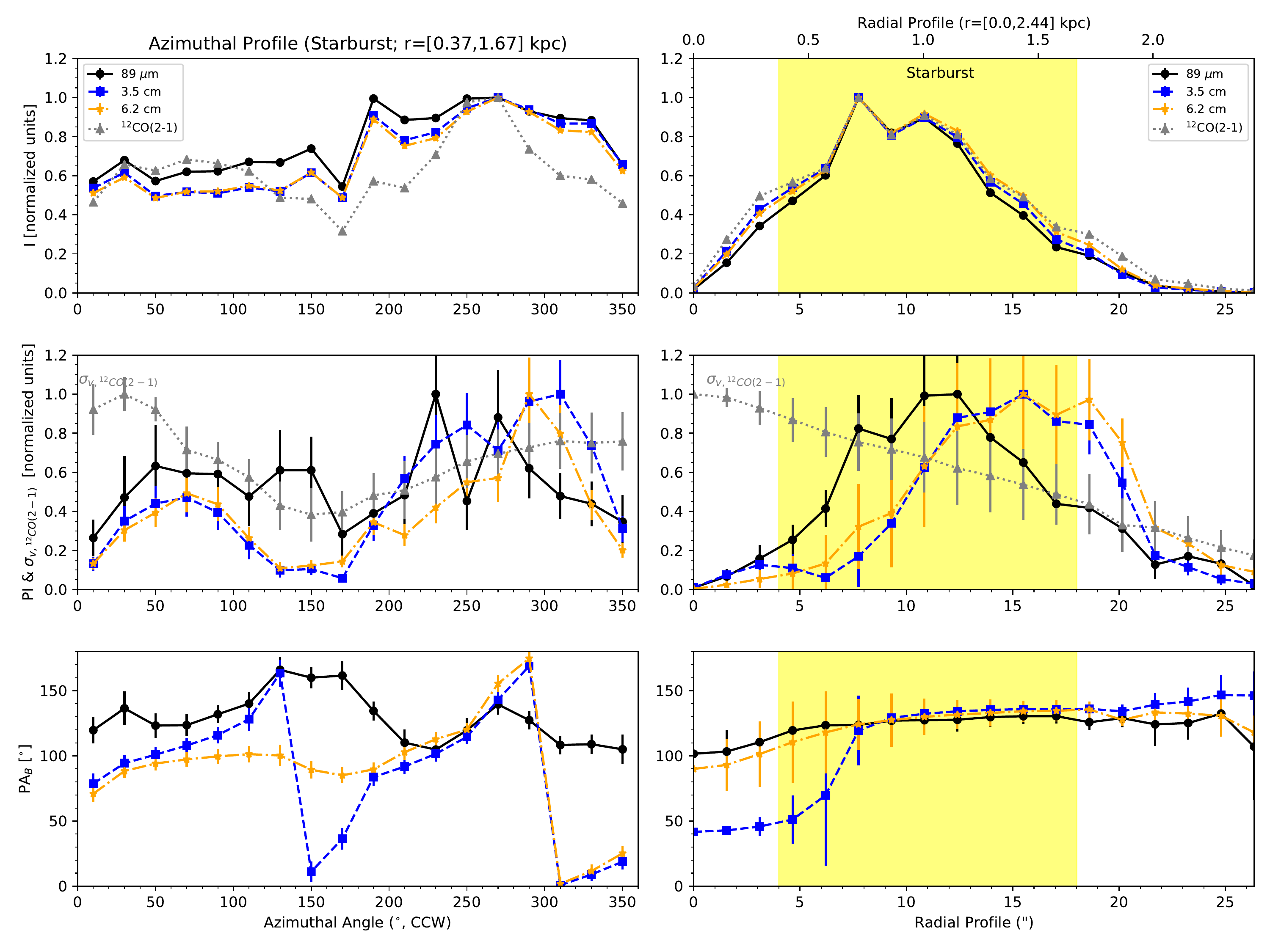}
\caption{Azimuthal (left column) and radial (right column) profiles of the total intensity (top), polarized intensity (middle) and B-field orientation (bottom) at 89 \um\ (black circle, solid line), 3.5 cm (blue square, dashed line), 6.2 cm (orange star, dotted-dashed line), and $^{12}$CO(2-1) (grey triangle, dotted line). Total and polarized intensities have been normalized to unity at the peak of their profiles. \lineco\ velocity dispersion, $\sigma_{v,^{12}\mathrm{CO(2-1)}}$, (grey triangle, dotted line) is shown in the middle panels. Azimuthal angles are counted east of north (counterclockwise direction, CCW). Radial profiles span the physical scales of $[0-2.44]$ kpc radius from the AGN position at a resolution of a quarter beam, $1.95\arcsec$. The physical limits of the starburst region, $[0.46,1.67]$ kpc, used for the azimuthal profiles is shown as yellow shaded area and red thin solid lines in Fig. \ref{fig:fig1}. 
\label{fig:fig3}}
\epsscale{2.}
\end{figure*}
%%%%%%%%%%%%%%%%%

Figure \ref{fig:fig0} shows the B-field orientation (background streamlines) and direction (black and white streamlines) using the radio polarimetric observations presented in Section \ref{subsec:AD}, and $89$ \um\ polarimetric observations (yellow lines). The B-field direction is estimated using Faraday rotation measurements shown in Section \ref{subsec:RM}. The colorscale shows the ultraviolet emission using the \textit{Hubble Space Telescope} (\textit{HST})/WFC3/F438W as a proxy for the star-forming regions in the starburst ring \citep[i.e.][]{Prieto2019}. This figure illustrates the main results from this work. Figure \ref{fig:fig1} shows the total and polarized intensity images at $89$ \um\ for the central $50\times50$ arcsec$^{2}$ ($4.63\times4.63$ kpc$^{2}$) region of NGC~1097. Each image has the polarization angles rotated by $90^{\circ}$ to show the orientation of the B-field.  Polarization measurements with $P/\sigma_{P} \ge 2$ are shown, where $\sigma_{P}$ is the uncertainty in the degree of polarizarion. In total intensity, the starburst ring is resolved at $89$ \um, which is consistent with previous observations using PACS/\textit{Herschel} by \citet{Sandstrom2010}. 

The polarized flux is mostly concentrated along the East and West regions and spatially coincident with the contact regions of the outer-bar (i.e. dust lane) with the starburst ring. Outside of the starburst ring, the B-field is mostly parallel to the outer-bar  orientation. Inside of the starburst ring, the B-field in the East region slightly curves toward the galaxy's center in a spiral-like shape. The Southern regions inside the starburst ring are unpolarized, although a change in the B-field morphology towards the central active nucleus is observed in the south-west region of the inner side of the starburst ring. The active nucleus is consistent with an unpolarized source.

\subsection{Archival data}\label{subsec:AD}

We use a set of archival data as part of the analysis in this paper. Specifically, we use the $3.5$ cm and $6.2$ cm radio polarimetric observations with an angular resolution of $6\arcsec$ by \citet{Beck2005}. The $22$ cm observations are strongly affected by Faraday rotation and Faraday depolarization; thus we do not use these images for our analysis. In addition to these, we use the \textit{HST}/WFC3/UVIS F438W as a proxy of the on-going star formation in the starburst ring \citep[i.e.][]{Prieto2019}. We use the \lineco\ observations from the Physics at High Angular resolution in Nearby GalaxieS (PHANGS) Survey\footnote{PHANGS: \url{https://sites.google.com/view/phangs/home}} taken with ALMA with a resolution of $1\farcs7$ by \citet{Leroy2021}. Radio and \lineco\ observations were smoothed using a Gaussian profile with a FWHM equal to the resolution of the HAWC+ observations and projected to the HAWC+ pixel grid (Fig. \ref{fig:fig1}). Specifically, the integrated emission line (moment 0) and velocity dispersion (moment 2) maps were smoothed and reprojected.

To compute the temperature and column density maps, we registered and binned $70-160$ \um~\textit{Herschel} observations taken with PACS to the same pixel scale, $1.95$\arcsec, of the HAWC+ observations. This approach ensures that images at all wavelengths have the same pixelscale and array dimensions. Then, for every pixel we fit an emissivity modified blackbody function with a constant dust emissivity index $\beta = -2.0$ \citep{Galametz2012}. We derived the molecular hydrogen optical depth as $N_{\mathrm{H2+HI}} = \tau/(k\mu m_{\mathrm{H}})$, with the dust opacity $k = 0.1$ cm$^{2}$ g$^{-1}$ at $250$ \um~\citep{H1983}, and the mean molecular weight per hydrogen atom $\mu = 2.8$. Temperature and column density values range from $[22-32]$ K and $\log(N_{H2+HI}) = [21-22]$ cm$^{-2}$  in agreement with \citet{Galametz2012}. Note that our maps (Fig. \ref{fig:fig6}) resolve the central starburst ring, which is unresolved in the maps by \citet{Galametz2012}.

%%%%%%%%%%%%%%%%%%%
%%%%% RESULTS %%%%%
%%%%%%%%%%%%%%%%%%%

\section{Thermal polarized emission and the multi-phase ISM}\label{sec:ANA}

%%%%%%%%%%%%%%%%%%
%%%% FIGURE 4 %%%%
%%%%%%%%%%%%%%%%%%
%% The "ht!" tells LaTeX to put the figure "here" first, at the "top" next
%% and to override the normal way of calculating a float position
\begin{figure*}[ht!]
\includegraphics[angle=0,scale=0.8]{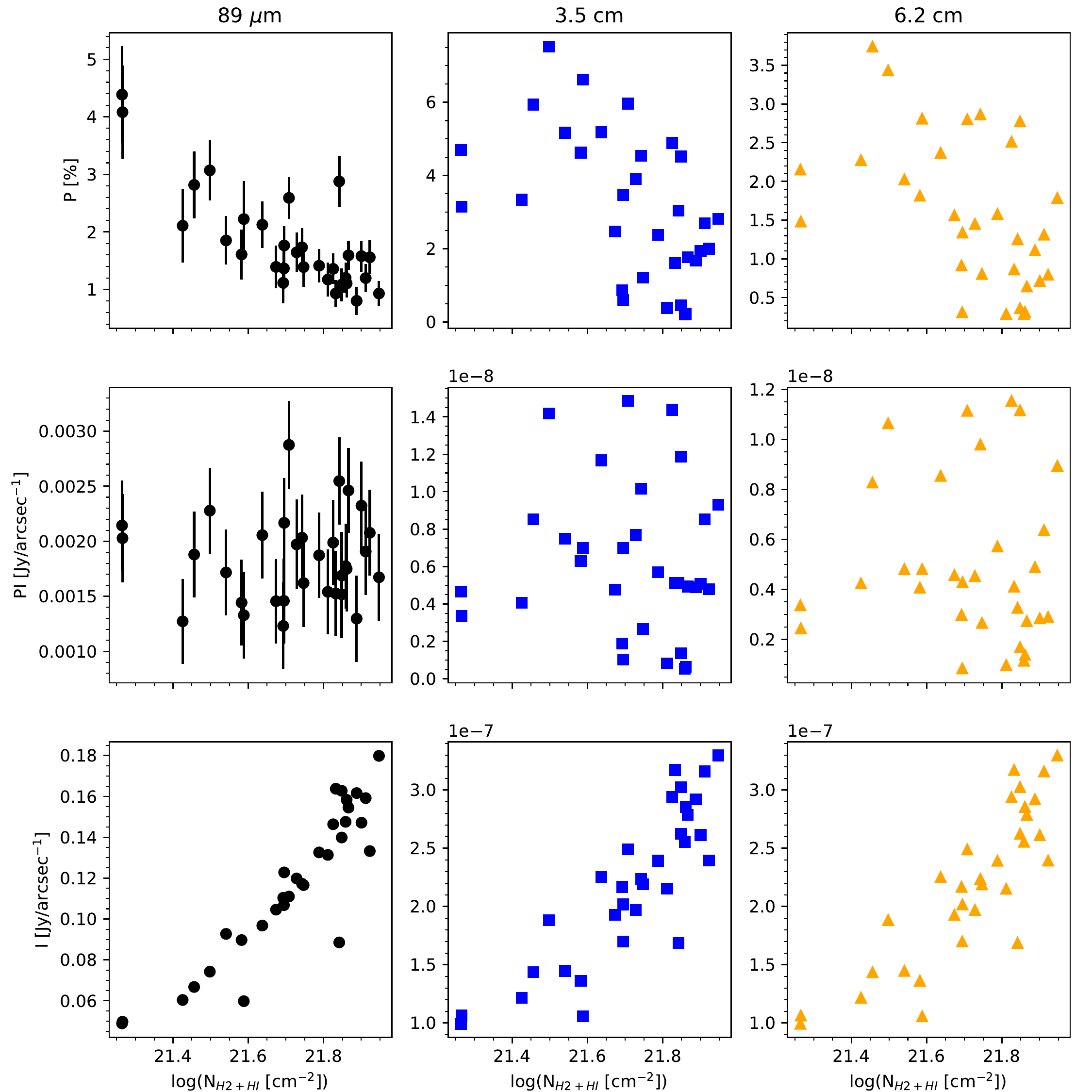}
\caption{Polarization fraction (top), polarized flux (middle), and total intensity (bottom) as function of the column density at $89$ \um~(left), $3.5$ cm (middle), and $6.2$ cm (right). Measurements with $p \le 30\%$, $SNR_I \ge 100$, and $SNR_p \ge 2$, and two measurements per beam are shown.
\label{fig:fig4}}
\epsscale{2.}
\end{figure*}
%%%%%%%%%%%%%%

 \subsection{Radial and azimuthal profiles}\label{subsec:RadAZ}

%%%%%%%%%%%%%%%%%%
%%%% FIGURE 5 %%%%
%%%%%%%%%%%%%%%%%
%% The "ht!" tells LaTeX to put the figure "here" first, at the "top" next
%% and to override the normal way of calculating a float position
\begin{figure*}[ht!]
\includegraphics[angle=0,scale=0.8]{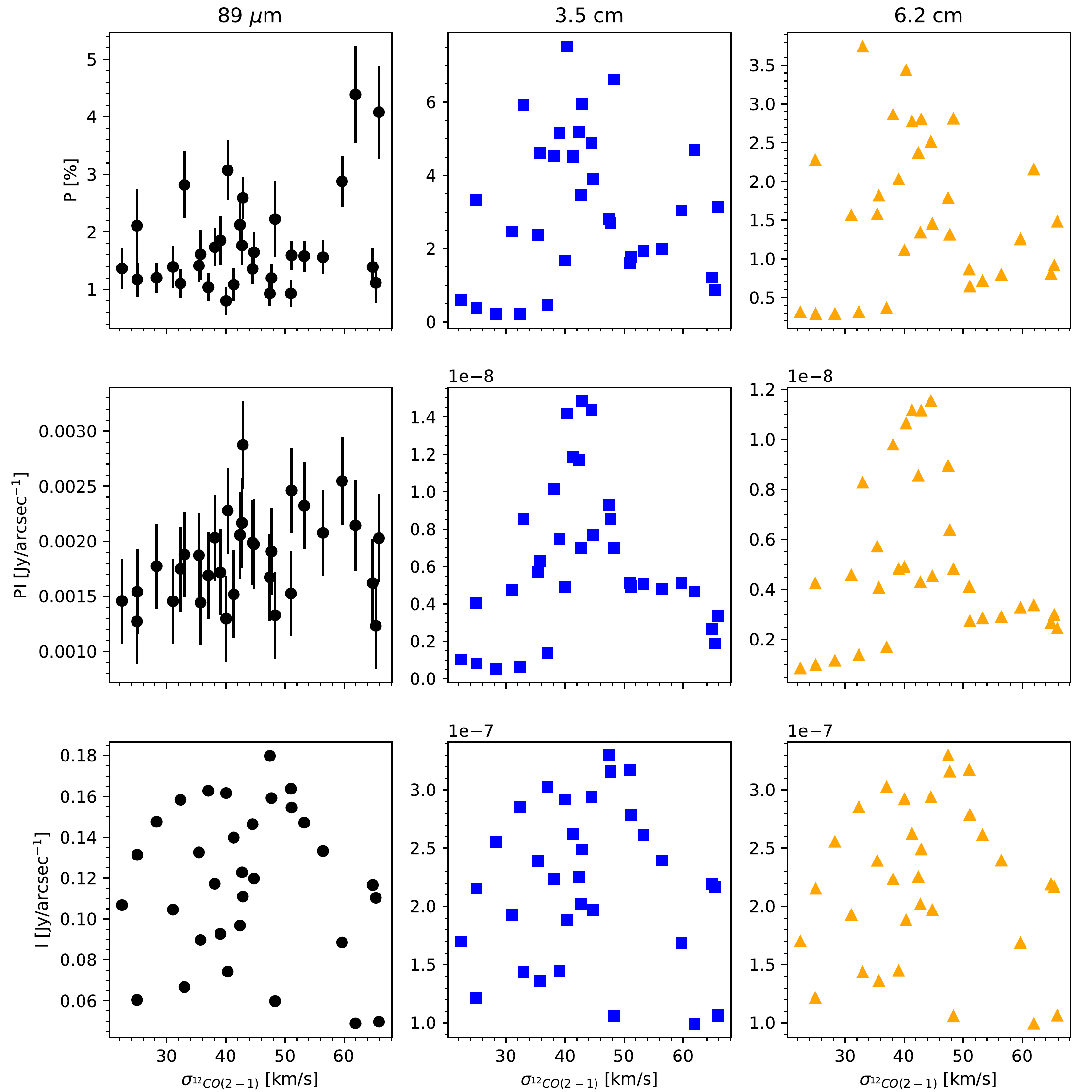}
\caption{Same as Fig. \ref{fig:fig4} but as a function of the velocity dispersion of the $^{12}CO(2-1)$ emission line.
\label{fig:fig5}}
\epsscale{2.}
\end{figure*}
%%%%%%%%%%%%%%%%%%
 
Figure \ref{fig:fig1} shows the comparison of the total and polarized intensities at $89$ \um, $3.5$ cm, and $6.2$ cm. This figure also shows the comparison of the FIR polarization with the total intensity (moment 0) of the \lineco\ emission line. The polarization measurements are overlaid on the total intensity. The orientation of the dust lane in the outer-bar (red solid lines) is parallel to the extended emission of the lowest contour of the \lineco\ integrated emission line shown in Figure \ref{fig:fig1}. This is only used for labeling purposes to identify the physical regions in NGC1097. Figure \ref{fig:fig2} shows a summary of the main features described hereafter.
 
We compute radial profiles in the range of $[0,2.44]$ kpc at a quarter beam resolution to show general trends. All radial profiles are estimated on the plane of the sky. We find that the total integrated fluxes at FIR, radio, and \lineco\ wavelengths are co-spatial along the distance from the galaxy's center (radial profile, Fig.\,\ref{fig:fig3}, top-right). However, our results show a spatial shift of the polarized flux and \lineco\ emission line across the starburst ring as a function of the distance to the galaxy's center. The \lineco\ integrated emission line has a peak in the inner $\sim7.5\arcsec$ ($\sim0.69$ kpc), while the  FIR polarized flux peaks in the central $\sim11\arcsec$ ($\sim1.02$ kpc), and the radio polarized flux peaks in the outer $\sim15\arcsec$ ($\sim1.39$ kpc) of the starburst ring. For both FIR and radio wavelengths, the polarized flux emission is mostly located at the contact regions of the outer-bar with the starburst ring in the East and West sides. The radio polarized emission shows extended emission in the North and South regions at a signal-to-noise ratio (SNR) of $3-6\sigma$.  

%%%%%%%%%%%%%%%%%%
%%%% FIGURE 6 %%%%
%%%%%%%%%%%%%%%%%%
%% The "ht!" tells LaTeX to put the figure "here" first, at the "top" next
%% and to override the normal way of calculating a float position
\begin{figure*}[ht!]
\includegraphics[angle=0,scale=0.39]{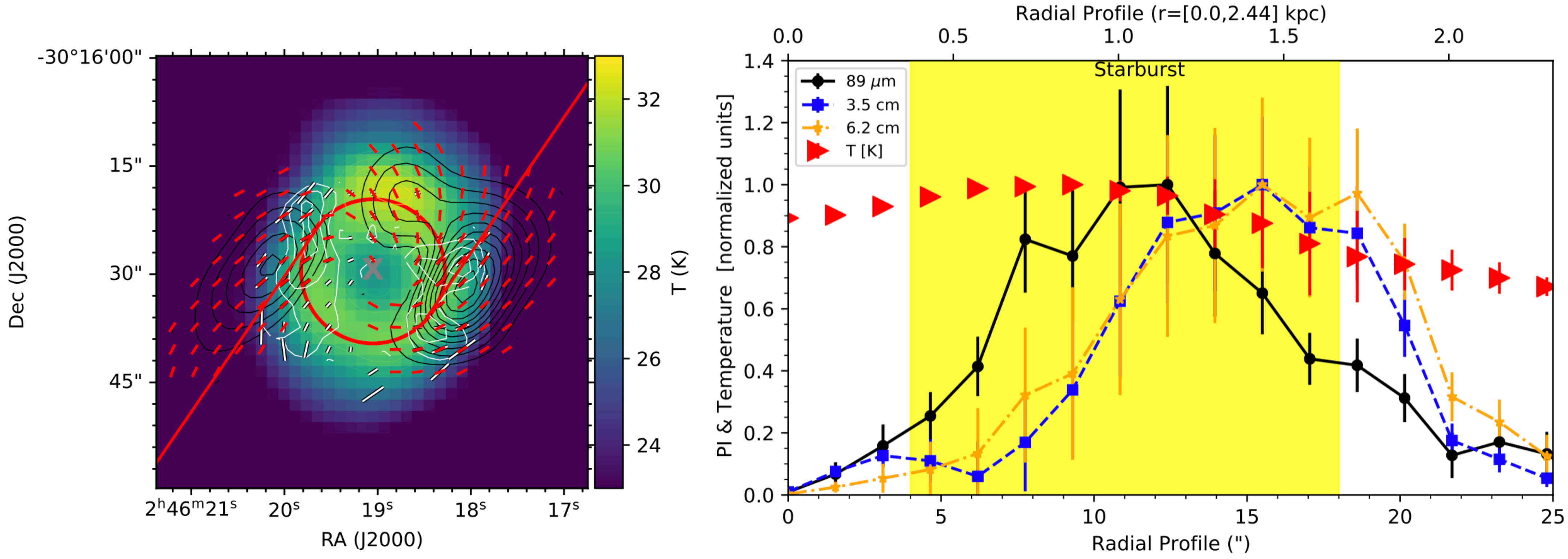}
\caption{\textit{Left:} Dust temperature map with overlaid B-field orientations at $89$ \um\ (white) and $3.5$ cm (red) and polarization fluxes at 89 \um~(white) and $3.5$ cm (black) as in Figure \ref{fig:fig3}. The central position of the starburst ring (red circle) and dust lanes (red solid lines) are shown. The AGN location (grey cross) is shown. \textit{Right:} Radial profiles of the polarized fluxes as in Figure \ref{fig:fig4} with the normalized temperature (red triangles) to the peak of the profile. The dust temperature is higher, $T_{89\,\mu \mathrm{m}} = 30.7\pm0.4$ K, at the position of the $89$ \um\ polarized flux than at the position of the $3.5$ and $6.2$ cm, $T_{3\,\mathrm{cm}} = 26.2\pm1.7$ K.
\label{fig:fig6}}
\epsscale{2.}
\end{figure*}
%%%%%%%%%%%%%%

We compute azimuthal profiles within an annulus of $[0.46-1.67]$ kpc (Fig. \ref{fig:fig1}) using angular widths of $20^{\circ}$ covering the starburst ring (Fig.\,\ref{fig:fig3}), which samples the azimuthal profiles at a half-beam resolution. All azimuthal profiles are estimated on the plane of the sky. The inner edge of the starburst ring was selected at the location of half-flux from the peak intensity at $89$ \um, and the outer edge at a level of $20$\% from the peak intensity at $89$~\um\ to account for the polarized flux at radio wavelengths (yellow region in Fig. \ref{fig:fig3}). We find that the total fluxes at the four different wavelengths are co-spatial around the starburst ring (azimuthal profile, Fig. \ref{fig:fig3} top-left panel). For the East side (AZ$\sim[50-150]^{\circ}$), the peak of FIR polarized flux is co-spatial with a gap of \lineco\ emission and the contact point with the starburst ring. For the Southwest side (AZ$\sim[220-280]^{\circ}$), the FIR polarized flux peaks at the edges of the extended $^{12}$CO(2-1) emission region. The radio and FIR polarized fluxes show similar azimuthal profiles, although the radio polarized flux is more extended in the south- and south-west sides of the starburst ring.

Regarding the B-field orientations, the outer ($r>10$\arcsec), and inner ($r<10$\arcsec) starburst ring need to be analyzed individually. We take a radius of $10\arcsec$ ($0.93$ kpc) as the middle point of the starburst ring (Fig. \ref{fig:fig1} and \ref{fig:fig2}). The B-field orientation is almost constant, $126\pm18^{\circ}$, in the FIR across the East and West sides of the starburst ring. In the outskirts of the starburst ring ($r>10$\arcsec), the B-fields on the East and West sides are mostly in the same orientation for both radio and FIR. The B-field orientations are parallel to the dust lane and encounters the starburst ring on the East and West sides. In the inner regions of the starburst ring ($r<10$\arcsec), the B-field orientations on the East and West regions twist towards the galaxy's center. This twist is more clear at radio than at FIR wavelengths (see Section \ref{subsec:LPD}). For the North and South regions, we find no individual statistically significant polarization measurements in the FIR. At radio wavelengths, the B-field has a spiral shape with an entry point towards the galaxy's center at $\sim30^{\circ}$, which is parallel to the inner-bar at a PA of $\sim28^{\circ}$ \citep{Quillen1995}. The B-field orientation at $3.5$ cm has a larger azimuthal variation than at $6.2$ cm (Fig. \ref{fig:fig3}-bottom-left), specifically, at $\sim150^{\circ}$, where the $3.5$ cm has an almost north-south ($0^{\circ}$) orientation. We study the behaviour of the B-field modes, pitch angles, and direction in Section \ref{sec:FIRvsRadio}. At the core, the active nucleus is consistent with an unpolarized source at FIR and radio wavelengths.

In summary, our results show that:

\begin{itemize}
\item The FIR polarized flux is co-spatial with the contact regions of the outer-bar with the starburst ring. The radio polarized flux shows an extended structure outside and within the starburst ring.
\item A spatial shift of the polarized fluxes in radio, FIR, and \lineco\ emission line across the starburst ring as a function of the distance to the AGN. The \lineco\ integrated emission line has a peak in the inner region, $\sim7.5\arcsec$ ($\sim0.69$ kpc), of the starburst ring, while the  FIR polarized flux peaks in the middle, $\sim11\arcsec$ ($\sim1.02$ kpc), and the radio polarized flux peaks in the outer edge, $\sim15\arcsec$ ($\sim1.39$ kpc), of the starburst ring.
\item In the FIR, the B-field orientation is mostly constant, $126\pm18^{\circ}$, on the East and West sides of the starburst ring. At radio, the B-field orientation show a spiral structure towards the AGN with entry points at $\sim30^{\circ}$.
\item The active nucleus is consistent with an unpolarized source at FIR and radio wavelengths.
\end{itemize}

\subsection{Gas kinematics, column density, and temperature as a function of polarization}\label{subsec:PIplots}

To study the effect of the ISM in the B-field and dust grain alignment efficiency traced at radio and FIR wavelengths, we use the column density and temperature. We also use the velocity dispersion of the molecular gas, \lineco, as a proxy of the turbulent kinetic energy. 

In terms of the column density (Fig.\,\ref{fig:fig4}), we find that the FIR and radio total intensities increase with column density. In the FIR, the polarization fraction decreases with increasing column density. However, there is no clear trend between the polarization and column density at radio wavelengths. Our results show no trends between the polarized flux and the column density at any of the three wavelengths. 

In terms of the molecular gas kinematics (Fig. \ref{fig:fig5}), the only clear trend is the almost constant $89$ \um\ polarization fraction with velocity dispersion of the molecular gas. For the rest of parameters, P and PI, there are no clear correlations at any of the three wavelengths within the starburst ring. From Fig. \ref{fig:fig3}-middle-right, from the inner to the outer edges of the starburst ring, the FIR and radio polarized fluxes increase as the velocity dispersion of the molecular gas decreases. But, within the starburst ring, our results show no trends between the polarized flux and the velocity dispersion at any of the three wavelengths (Fig. \ref{fig:fig5}). From the azimuthal profiles (Fig. \ref{fig:fig3}-middle-left), there is a small shift of $\ge20^{\circ}$ between the peaks of FIR and radio polarized fluxes and the velocity dispersion of the molecular gas at azimuthal angles of $0-50^{\circ}$ and $220-330^{\circ}$ (Fig. \ref{fig:fig3}-right). Figure \ref{fig:fig2} shows that the star-forming regions (\textit{HST/}WFC3/F438W) and the CO emission are not cospatial, which was also found by \citet{Hsieh2011}. The azimuthal profiles show that the velocity dispersion of the molecular gas is higher at the locations of star-forming region, i.e. AZ $\sim20-90^{\circ}$ and $250-360^{\circ}$.

In terms of the dust temperature (Fig. \ref{fig:fig6}), we find that the dust temperature is higher, $T_{89\mu \mathrm{m}} = 30.7\pm0.4$ K, at the peak position of the $89$ \um\ polarized flux than at the position of the $3.5$ and $6.2$ cm, $T_{3\,\mathrm{cm}} = 26.2\pm1.7$ K. The dust temperatures were estimated within the ranges of distances of $[6.2-12.4]\arcsec$ ($0.57-1.15$ kpc) at $89$ \um, and $[12.4-18.6]\arcsec$ ($1.15-1.72$ kpc) at radio wavelengths.

%%%%%%%%%%%%%%%%%%%%%%%%%%%%%%%%%
%%%%% RADIO vs FIR B-FIELDS %%%%%
%%%%%%%%%%%%%%%%%%%%%%%%%%%%%%%%%

\section{Magnetic field components at FIR and Radio wavelengths} \label{sec:FIRvsRadio}

Our observations have shown that both FIR and radio wavelengths may trace different B-field components of the multi-phase ISM. To quantify the B-field morphology for both wavelength regimes, we perform three analyses: 1) estimation of the pitch angles as a function of the azimuthal angle, 2) linear polarimetric decomposition within the starburst ring, and 3) Faraday rotation measurements (RMs). The first analysis allows us to quantify the variation of the orientation of the B-field across the starburst ring. The second analysis allows us to quantify the dominant modes of the B-field orientations within the starburst ring. The third analysis allows us to estimate the LOS magnetic field sign in the central $1$ kpc.

\subsection{Pitch angles of the B-field}\label{subsec:pitchangle}

\begin{figure}[ht!]
\includegraphics[angle=0,scale=0.62]{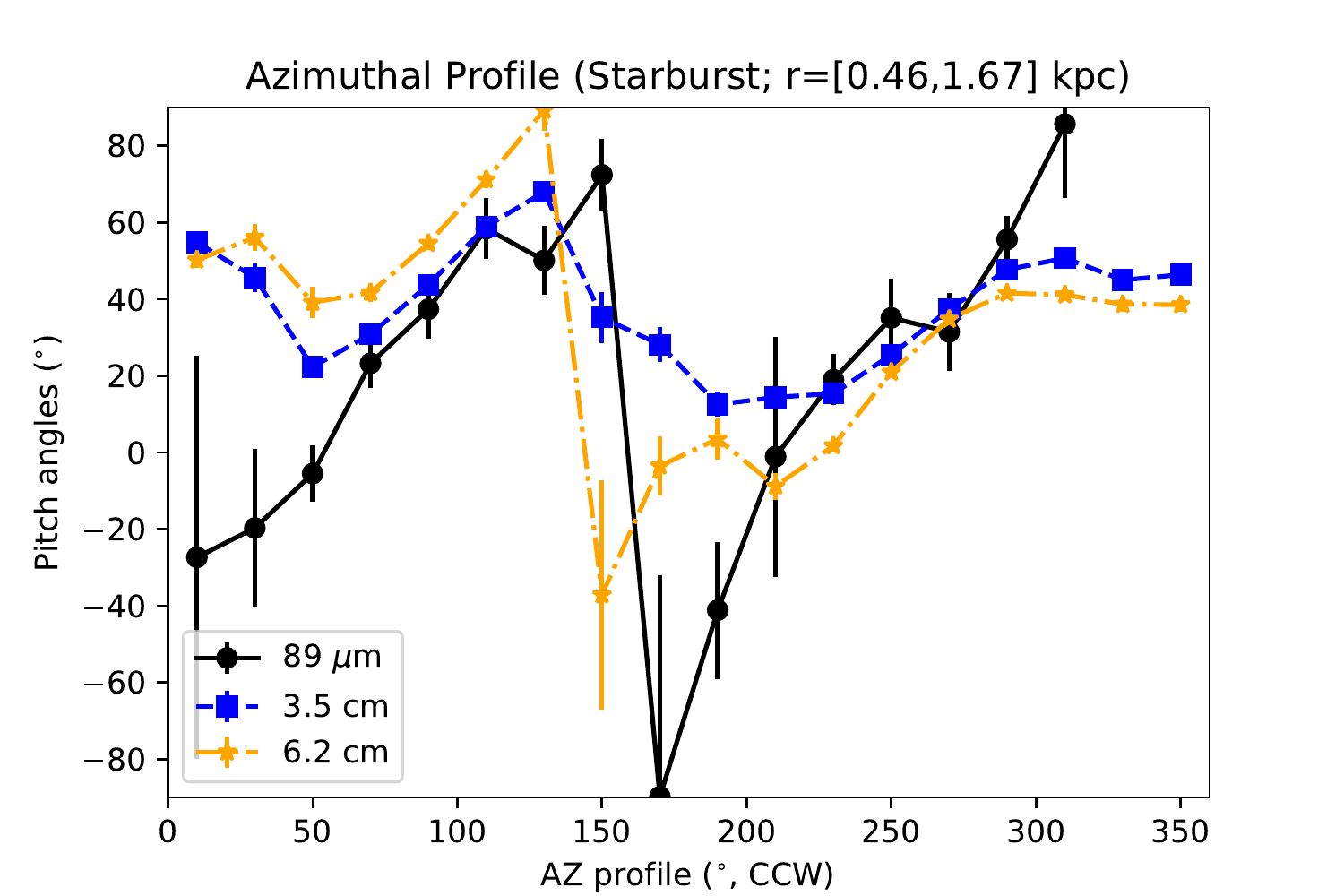}
\caption{Pitch angles of the B-field morphology at $89$ \um\ (black), $3.5$ cm (blue), and $6.2$ cm (orange) as a function of the azimuthal angle within an annulus in the range of $[0.74-1.67]$ kpc. An angular sector width of $20^{\circ}$ was used.
\label{fig:fig7}}
\epsscale{2.}
\end{figure}

In addition to the measured B-field orientations as a function of azimuth within the starburst ring and radial distance to the galaxy's center (Fig. \ref{fig:fig3}), we compute the pitch angles of the B-field as a function of the azimuthal angle within the starburst ring.  Figure \ref{fig:fig7} shows the pitch angles at $89$ \um, $3.5$ cm, and $6.2$ cm as a function of the azimuthal angle within the $[8-18]\arcsec$ ([$0.74-1.67$] kpc) starburst ring. Note that the annulus is narrower than that used in Section \ref{subsec:RadAZ} to optimize the contribution of the polarized flux at radio wavelengths within the starburst ring. The minimum at $8$\arcsec\ in the radial profile of the polarized flux at radio wavelengths was used as the inner radius. We compute the pitch angles within sectors of angular width of $20^{\circ}$, where the 1-$\sigma$ uncertainty for Stokes $Q$ and $U$ are estimated to be $0.39$ mJy arcsec$^{-1}$, $0.6$~mJy beam$^{-1}$, and  $1.0$~mJy beam$^{-1}$ at $89$ \um, $3.5$ cm, and $6.2$ cm, respectively.

At FIR wavelengths, the pitch angle shows an almost linear variation with the azimuthal angle, which is the signature of a single B-field mode with an almost constant orientation. This result is in agreement with Fig. \ref{fig:fig3}, where we show that there is an almost constant B-field orientation, $126\pm18^{\circ}$, with azimuth. At radio wavelengths, the pitch angles show a significant variation with azimuth within the ranges of $[68,12]^{\circ}$ and $[89,-37]^{\circ}$ at $3.5$ and $6.2$ cm, respectively. This behaviour is indicative of the contribution of several B-field modes. 

%%%%%%%%%%%%%%%%%%
%%%% FIGURE 7 %%%%
%%%%%%%%%%%%%%%%%%
%%%%%%%%%%%%%%

%%%%%%%%%%%%%%%%%%
%%%% FIGURE 8 %%%%
%%%%%%%%%%%%%%%%%%
\begin{figure*}[ht!]
\includegraphics[angle=0,scale=0.55]{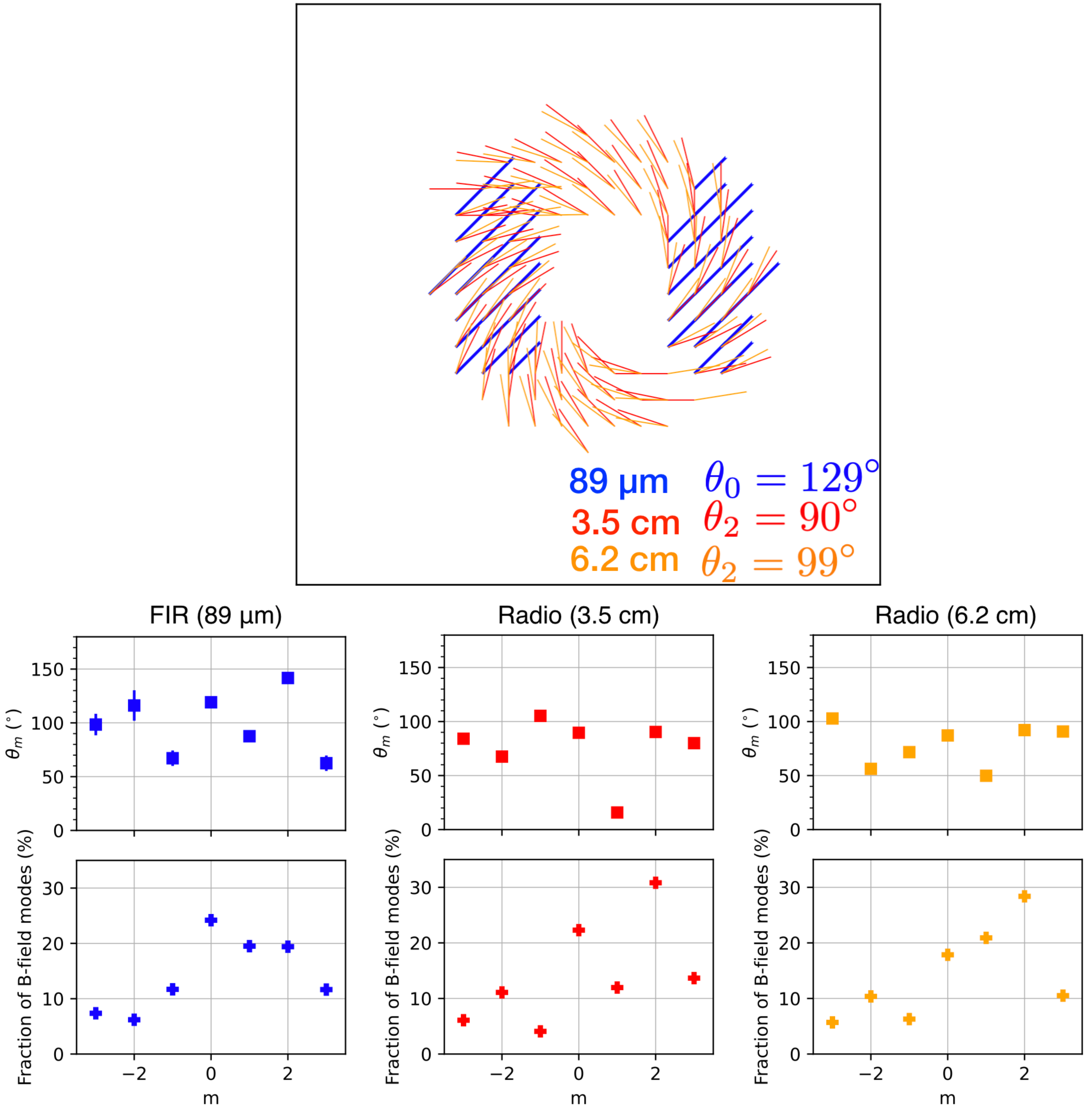}
\caption{B-field orientations (top) from the linear polarimetric decomposition within the annulus of the starburst ring of NGC~1097 at $89$ \um\ (blue), $3.5$ cm (red), and $6.2$ cm (orange). The B-field orientations were generated using the dominant amplitudes per wavelength. The fractional contribution of the amplitude, $\beta_{m}$, and phase, $\theta_{m}$, for the range of azimuthal modes, $m=[-3,3]$, of the complex polarization brightness distribution are shown at the bottom at $89$ \um\ (left), $3.5$ cm (middle), and $6.2$ cm (right). 89 \um\ is dominated by power in the $m=0$, while $m=2$ dominates at $3$ and $6$ cm at a level of $25-30$\% (bottom panels). For each wavelength, other modes are also present at a level of $\sim20$\%. Note the similarities with the polarization map shown in Fig. \ref{fig:fig2}.  
\label{fig:fig8}}
\epsscale{2.}
\end{figure*}
%%%%%%%%%%%%%%

\subsection{Azimuthal B-field modes}\label{subsec:LPD}

To quantify the B-field modes within the starburst ring, we use the linear polarimetric decomposition introduced by \citet{Palumbo2020}. This method has been successfully applied to analyze the B-field morphology around the black hole of M87 \citep{EHTVII,EHTVIII}, which provides a model-independent approach to estimate the dominant B-field mode of a system.  We here describe and apply this method to the starburst ring of NGC~1097.

 The linear polarization is decomposed in a complex polarization field, $P(\rho,\phi) = Q(\rho,\phi) + iU(\rho,\phi)$, where $Q$ and $U$ are the Stokes parameters. This formulation allows the decomposition in azimuthal modes, $m$, with an amplitude, $\beta_{m}$, given by
 
 \begin{equation}
 \beta_{m} = \frac{1}{I_{ann}}  \int_{r_{min}}^{r_{max}}\int_{0}^{2\pi} P(\rho,\phi) e^{-im\phi} \rho~d\phi d\rho
 \label{eq:betam}
 \end{equation}
 \noindent
 where $I_{ann}$ is the integrated Stokes I of an  annulus within $r_{min}$ and $r_{max}$, such that

%%%%%%%%%%%%%%%%%
%%%% TABLE 2 %%%%
%%%%%%%%%%%%%%%%%
\begin{deluxetable*}{lcccc}[ht!]
\tablecaption{Parameters of the dominant azimuthal B-field mode from the linear polarimetric decomposition (Fig. \ref{fig:fig8}).\label{tab:table2}}
\tablecolumns{6}
\tablewidth{0pt}
\tablehead{
\colhead{Parameter} & \colhead{Symbol}	 & 	\colhead{89 \um}	&	\colhead{3.5 cm}	&
\colhead{6.2 cm}  
}
\startdata
Mode		&	$m$				&	$0$				&	$2$				&	$2$	\\
Amplitude		&	$|\beta_{m}|$		&	$0.017\pm0.001$	&	$0.0367\pm0.0005$	&	$0.0280\pm0.0005$	\\
Phase Angle 	&	$\theta_{m}$	&	$129\pm3^{\circ}$	&	$90\pm1^{\circ}$	&	$99\pm1^{\circ}$	\\
Fractional contribution &				&	$24$\% 			&	$30$\%			&	$27$\%			
\enddata
\end{deluxetable*}
%%%%%%%%%%%%%%%%%%%%
 
 \begin{equation}
 I_{ann} = \int_{r_{min}}^{r_{max}}\int_{0}^{2\pi} I(\rho,\phi) \rho~d\phi d\rho
 \label{eq:Iann}
 \end{equation}

$\beta_{m}$ is a dimensionless complex number, which corresponds to the amount of coherent power in the $m$-th mode. The complex phase of the amplitude $\beta_{m}$ component is $\theta_{m}$ and represents the dominant orientation of the B-field mode within the range of $[0,180)^{\circ}$. Note the change from $[-90,90]^{\circ}$ by \citet{Palumbo2020} and the new range here to be consistent with the nomenclature of position angles throughput this manuscript. The phase shows the averaged pointwise rotation of the B-field orientation with $\phi =0$ in the north direction and positive amount along the counterclockwise direction. Under this formulation, a mode $m=0$ shows a constant B-field orientation at an angle given by $\theta_{0}$, and a mode $m=1$ shows a spiral B-field shape with a dominant orientation of $\theta_{1}$. The $m=2$ mode describes a rotationally invariant polarization pattern that is directly analogous to the $E$- and $B$-mode decomposition that is widely used in analysis of the cosmic microwave background \citep{Kamionkowski1997, Seljak1997, Zaldarriaga2001}, and increasingly used to characterize polarized dust in the Galactic ISM \citep[e.g.,][]{PlanckCollaboration2020, Clark2021}. Higher modes, $m\ge3$, are complex Fourier compositions of the complex azimuthal modes with a dominant orientation given by $\theta_{m}$. An example of these modes is shown in Fig.\,1 by \citet{Palumbo2020}. Note that these modes represent the overall B-field orientation within an annulus at a given radius and should not be confused with the mode-solutions of the mean-field dynamo equations, which also provide the B-field direction \citep{Beck1996}.

We perform the linear polarization decomposition analysis in the starburst ring covering an annulus within $r_{min} =0.46$ kpc and $r_{max} = 1.67$ kpc (Fig.\,\ref{fig:fig3}). This range is based on the total intensity image at $89$ \um\ from the analysis performed in Section \ref{sec:ANA}. The central coordinate is located at the position of the peak of total intensity at \lineco\ emission line (Fig. \ref{fig:fig2}). We moved this central coordinate within $2$ pixels and found no changes in the final results of the linear polarization decomposition analysis. We explore the azimuthal modes within the range of $[-3,3]$. Note that the estimation of the amplitudes and phases are independent of the selected range of modes. Uncertainties are measured as follows. We compute a data cube of randomly distributed series of 5000 samples following a Gaussian distribution with a mean equal to the measured values of the Stokes IQU images and 1-$\sigma$ equal to the standard deviation within the background region of the Stokes IQU images. A value of the amplitude and phase angle are estimated for each sample of the data cube. Finally, for each mode, $m$, the median and standard deviation are computed using the entire sample.

Fig.\,\ref{fig:fig8} shows the amplitudes, $|\beta_{m}|$, phases, $\theta_{m}$, and their fractional contributions for each of the modes in the range of $m=[-3,3]$. The measured values of the dominant mode are shown in Table \ref{tab:table2}. We find that the B-field morphologies at FIR and radio have different configurations. At $89$ \um, the B-field morphology is dominated by power in the $m=0$ with $\theta_{0} = 129\pm3^{\circ}$ at a level of $24$\%. We find that the B-field at $89$~\um\ also has more complex modes, $m=1$ and $m=2$, which each contribute at a $\sim20$\% level.

At $3.5$ and $6.2$ cm, the dominant power is found to be at $m=2$ with $\theta_{2} = 90^{\circ}$ and $99^{\circ}$ at a level of $30$\% and $27$\%, respectively. The B-field at $3.5$ cm also has a $m=0$ at a $\sim22$\% level, while the B-field at $6.2$ cm has $m=1$ and $m=0$, which each contribute at a $\sim20$\% level. 

Fig.\,\ref{fig:fig8}-top shows the B-field orientations from the dominant configurations at each wavelength. Note the similarities of this analysis with that from the observed B-field morphology shown in Fig. \ref{fig:fig2}. For display purposes, at $89$ \um, only the B-field orientations along the contact regions of the outer-bar with the starburst ring are shown. These B-field orientations represent the statistically significant polarization measurements, i.e. $P/\sigma_P \ge2$, from our observations. The inferred B-field orientation is constant within the starburst ring. This result is as expected given that the measured polarized flux dominates in the contact regions of the outer-bar with the starburst ring. At radio, the B-field is consistent with a spiral-like shape with a dominant $m=2$ with the spiral more tightly wound at $6.2$ cm ($\theta_{2} = 99^{\circ}$) than at $3.5$ cm ($\theta_{2} = 90^{\circ}$).

\subsection{The sign of the B-field}\label{subsec:RM}

%%%%%%%%%%%%%%%%%%
%%%% FIGURE 9 %%%%
%%%%%%%%%%%%%%%%%%

\begin{figure}[ht!]
\includegraphics[angle=0,scale=0.6]{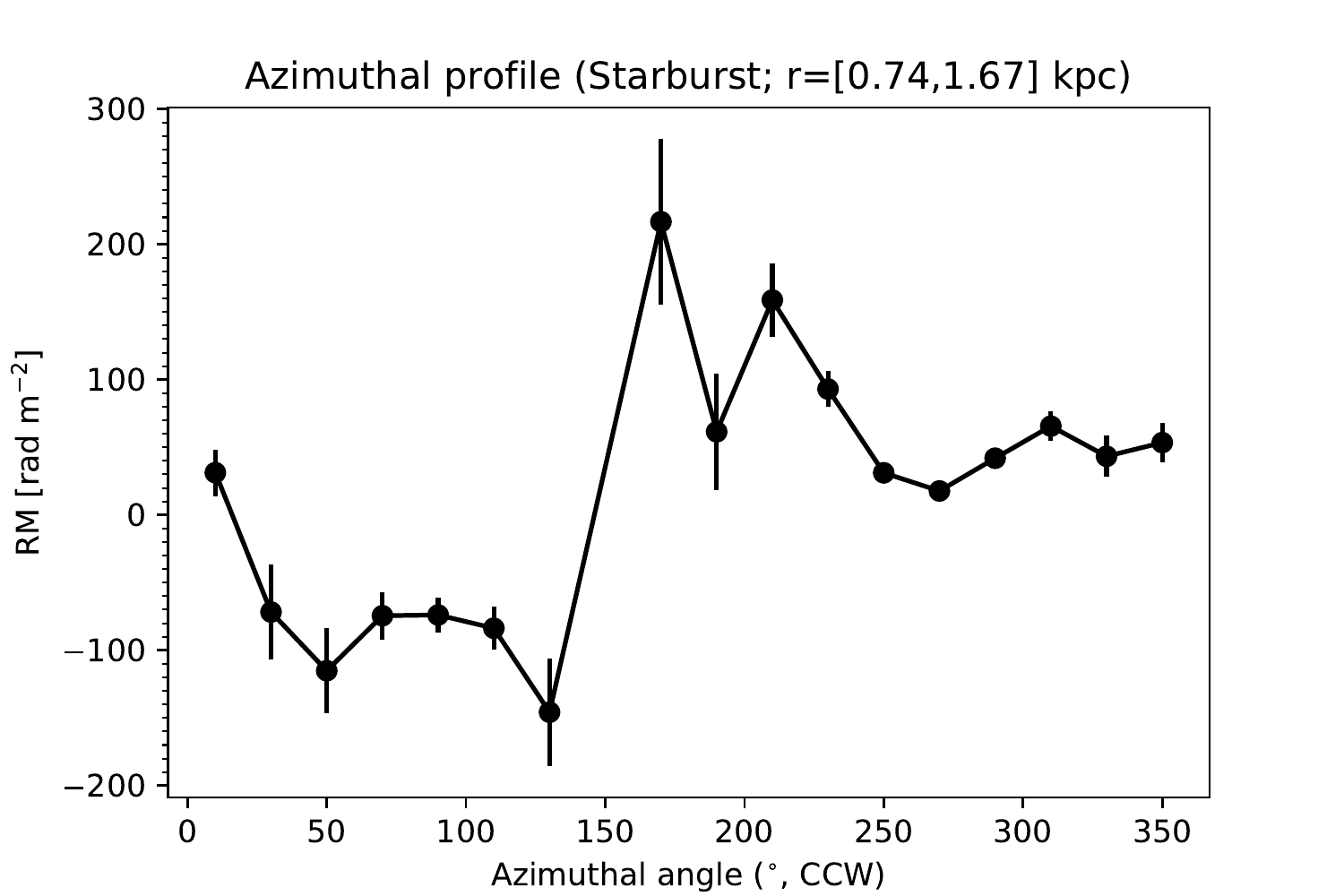}
\caption{The variation of the Faraday rotation measures (RMs) with azimuthal angle (counted counterclockwise from the North), computed using the $3.5$ and $6.2$\,cm data at $7\farcs8$ resolution (Section \ref{subsec:AD}). The RMs have two regimes: negative in the East [$20-120]^{\circ}$ and positive in the West [$160-360]^{\circ}$, which may indicate that the B-field in the contact regions has different signs. 
\label{fig:fig9}}
\epsscale{2.}
\end{figure}
%%%%%%%%%%%%%%

The radio data at $3.5$ and $6.2$\,cm allow us to compute Faraday rotation measures (RMs) and measure the sign of the ordered B-fields in the starburst ring. Fig.\,\ref{fig:fig9} shows the azimuthal variation of RMs averaged in $20^{\circ}$-wide sectors of an annulus between $8$\arcsec\ and $18$\arcsec\ [$0.74-1.67$] kpc radius. The averaging was performed in the Stokes Q and U data at the two frequencies to derive average polarization angles and their difference, as described in \citet{Beck2005}. RMs obtained between only two frequencies suffer from ambiguity, $\pm\,n \times 1.232$\,rad\,m$^{-2}$ in our case, which can be disregarded for the small RMs measured in NGC\,1097 \citep[Fig.\,11a in][]{Beck2005}, where $n$ is the number of rotations. The sector width corresponds to about half the beam size, so that the points are not independent. The 1-$\sigma$ uncertainty in RM was computed from the rms noise values in the images of Stokes Q and U via error propagation.

The measured RMs are mostly negative in the eastern half ($<RM>\,=-82 \pm 20$\,rad\,m$^{-2}$) and positive in the western half ($<RM>\,=+41 \pm 11$\,rad\,m$^{-2}$), similar to the values measured in the bar on larger scales \citep{Beck2005}. Negative (positive) RMs in the East (West) tell us that the LOS component of the B-field points away from us (towards us), i.e. the LOS of the observer. 

%%%%%%%%%%%%%%%%%%%%%
%%%% CONCLUSIONS %%%%
%%%%%%%%%%%%%%%%%%%%%

\section{Discussion}\label{sec:DIS}

\subsection{Origin of the thermal polarization}\label{subsec:Polori}

From Figure \ref{fig:fig2} and Section \ref{sec:ANA}, we find that a) the peak of $89$ \um\ polarized flux, $\sim11\arcsec$ radius, is slightly offset from the peak of \lineco\ at $\sim7.5\arcsec$ radius, b) the CO peaks do not coincide with the regions of star formation in the starburst ring, and c) the peaks of $89$ \um\ polarized flux do not coincide with the star forming regions. \citet{Hsieh2011} also found that the CO and star-forming regions are not spatially coincident in the starburst ring. A comparison between the total intensity maps (moment 0) of several tracers (HCN, HCO$^{+}$, C$_{2}$H, CS, H$^{13}$CN, H$^{13}$CO$^{+}$, HNCO, SiO, and HC$_{3}$N) using ALMA by \citet{Martin2015} and our FIR polarized flux shows that there is no spatial correspondence between the FIR polarized flux and any of these gas tracers in the starburst ring. If the dust grains are heated by the nearby star forming regions (Fig. \ref{fig:fig2}), one may have expected a patchy distribution of $89$ \um\ polarized flux across the starburst ring. Under this hypothesis, polarized flux at the locations of star-forming regions in the Northwest (AZ $\sim335^{\circ}$) and Southeast (AZ $\sim150^{\circ}$) should have been detected. These results show that the star forming regions are unpolarized, and that the FIR polarization in the starburst ring of NGC~1097 is not directly related to the star-forming regions. In addition, the peak of the 89 \um\ polarized flux is found to be colocated with the peak of dust temperature within the central $\sim1$ kpc (Figure \ref{fig:fig6}), and with an almost constant B-field orientation of $126\pm18^{\circ}$ (Section \ref{subsec:RadAZ}). We conclude that the 89 \um~polarized flux may arise from a compressed B-field in the warmest region at the contact points between the outer-bar and the starburst ring. 

The decrease in the FIR polarization with column density (Section \ref{subsec:PIplots}) may be due to 1) variations of the dust grain alignment efficiency, 2) variations of the B-field orientation in the plane of the sky due to an increase of gas turbulence, and/or 3) complex kinematics within the beam  \citep{LR2018a,Chuss2019,King2019,ELR2021,Borlaff2021}. A decrease of dust grain alignment efficiency may occur towards star-forming regions, which can cause a decrease of polarization with increasing column density. This effect may be due to a) collision dumping effects, which narrow the distribution of dust grain sizes, producing a decrease of $P$ with increasing $N_{H}$ or $I$ \citep[i.e.][]{Hoang2021}, and/or b) tangled B-fields along the LOS due to the star forming regions and an increase of the gas turbulence. The FIR polarized flux is colocated with a region where the B-field is compressed due to the shocked gas with a high B-field strength of $\sim60~\mu$G \citep{Beck1999,Beck2005}. The unpolarized FIR regions are found to be colocated with the star-forming regions in the starburst ring (Figure \ref{fig:fig2}). Star forming regions have tangled B-fields and larger turbulent gas than the contact points between the outer-bar and the starburst ring. Thus, the most likely scenario is that tangled B-fields along the LOS may explain the decrease of FIR polarization with column density. This is also a common feature found in thermal polarized emission in our Galaxy by \textit{Planck} polarimetric observations \citep{PlanckXX}. We suggest that the decrease of FIR polarization with increasing column density may be due to an increment of the small-scale turbulent fields (i.e. tangled B-fields) arising from the star forming regions at smaller scales ($<0.72$ kpc) than the beam of our observations. 

\subsection{The B-fields in the central 1 kpc}\label{subsec:Bori}

The spiral pattern seen in Fig.\,\ref{fig:fig8} (top panel) may suggest the existence of a regular axisymmetric spiral B-field, as found in several spiral galaxies via a sinusoidal variation of RM within the (inclined) galaxy plane \citep[table 5 in][]{Beck2019}. Such B-fields are regarded as evidence for the action of a large-scale dynamo.\footnote{The axisymmetric spiral field has a constant pitch angle in the galaxy's plane and is the lowest-mode solution of the mean-field dynamo equation \citep{Beck1996}. This dynamo mode is different from the mode as defined in Eq.~\ref{eq:betam}.} However, no such pattern is seen in the starburst ring of NGC\,1097 (Fig.\,\ref{fig:fig9}). Based on our results in Section \ref{subsec:RM} and knowing that the western side is nearer to us \citep[i.e.][]{Hsieh2011,PF2014}, we estimate that the radial component of the B-field along the contact regions between the starburst ring and the outer-bar is pointing inwards (i.e. towards the galaxy’s center) on both sides.

The existence of an additional axisymmetric B-field (or regular B-fields of higher symmetry) in the starburst ring cannot be excluded with the present radio data. New observations with higher resolution are needed. However, higher B-field modes were estimated using the linear polarimetric decomposition shown in Section \ref{subsec:LPD}. We find that there are major contributions of other B-field modes ($m=1$ and $2$) at all wavelengths. This result suggests that the measured B-field orientation at any given wavelength is a superposition of several B-field modes tracing both the diffuse and dense gas in the central $1$ kpc of NGC~1097. The starburst ring of NGC~1097 has a non-axysymmetric B-field. 

Section \ref{subsec:RadAZ} shows that the B-field in the dense ISM, traced by FIR, is dominated by a constant B-field orientation. This constant B-field is spatially coincident with the contact regions of the dust lane and the starburst ring, which we found to be cospatial with a peak of dust temperature. Radio polarized flux traces a cooler gas (i.e. diffuse ISM) in the outer-bar and it is dominated by a spiral-like shape outside and within the starburst ring towards the active nucleus (Section \ref{subsec:Bori}). The spiral B-field shows the signature of a galactic dynamo taking action outside and inside the starburst ring. Thus, the B-field in the diffuse ISM may be dragging material from the host galaxy to the galaxy's center, while the B-field in the dense ISM is compressed with a constant B-field orientation at the contact regions. 

\subsection{Physical scenario}\label{subsec:Scenario}

The B-field inferred from radio polarimetric observations is similar to the gas kinematics in a bar potential \citep{Beck2002, Beck2005}, while the B-field inferred from FIR polarimetric observations resembles the flow in a shock. These results show that a shock occurs in the dense and molecular cold gas, which compresses the B-field in the contact regions of the galaxy bar with the starburst ring observed at FIR wavelengths. The gas flow of the diffuse gas may be shearing the B-field observed at radio wavelengths rather than fully compressing it. Under this scenario, the B-field traced by radio contains several modes representing the compressing and shearing mechanisms. The B-field traced by FIR wavelength is dominated by a mode representing the compressed B-field. In addition, our measurement of the Faraday RM indicates that the B-field direction is pointing inwards towards the galaxy's center. 

As mentioned in the introduction, the composition of several B-field modes is a signature of a non-axisymmetric potential in barred galaxies \citep{Moss1998}. Two-dimensional MHD simulations show that shock waves at $\sim1$ kpc from the nucleus can be induced by magnetic stress \citep{Kim2012}.  Magnetic stress helps to remove angular momentum of the gas at the shocks, deflecting the gas flows to form a ring closer to the active nucleus that those from HD simulations. \citet{PF2014} observed bar-induced gas inflows in the diffuse ionized gas and a smaller ring that those predicted by their HD simulations. These authors suggested that the starburst ring may have formed at the Lindbland resonance radius and then migrated towards inner radii. Note the similarity between the spiral-like structure of the non-circular motions from kinematic data of several gas tracers (NII, HCN(4-3)) from kpc to pc scales \citep{vdV2010,Fathi2013}, and the B-field morphology at the three wavelengths presented by our work. Our results show that the B-field direction is pointing inwards, the position of the shocks in the contact points of the galaxy bar with the starburst ring is at $\sim1$ kpc, and the B-field is a composition of high modes (i.e. circular and spiral-like structures). All these results support the MHD framework \citep[i.e.][]{Moss1998,Kim2012} to explain the gas inflow towards the galaxy's center.

We conclude that 89 \um~traces the B-field in the dense gas of the starburst ring, which has been compressed by a shock caused by the galactic bar potential located at the contact regions of the outer-bar with the starburst ring. Radio wavelengths trace a spiral B-field in the diffuse ionized gas, which may be dominated by a non-axisymmetric perturbation of the gravitational potential in the galaxy bar \citep[i.e.][]{CL1994,Moss1998}.

The results presented here suggest a scenario where the molecular gas has been compressed along with the B-field in the dust lane and formed clouds. The B-field is then deflected into the starburst ring. The starburst ring is magnetically critical, which results in inefficient high-mass star formation \citep{Tabatabaei2018}. The B-field orientation is preserved in the contact regions due to shocked gas produced by the bar-driven gas towards the galaxy's center. This B-field then may be deflected and dragging the dense gas towards the orbit of the starburst ring. On the other hand, the radio emission from the relativistic electrons mixed in with the diffuse ionized ISM responds to shear flows, which drags the diffuse gas via a spiral B-field towards the active nucleus.  Here, we argued that the gas flow follows the B-field morphology towards the active nucleus. 

\section{Conclusions}\label{sec:CON}

We reported $89$ \um\ polarization observations using HAWC+/SOFIA of the circumnuclear starburst ring at the central $1$ kpc of the nearby barred galaxy NGC~1097.  We inferred the B-field orientation in the dense ISM using the thermal polarization of magnetically aligned dust grains. Using the linear polarization decomposition presented by the EHT polarimetric observations of M87 \citep{EHTVII,EHTVIII}, we found that the B-field morphologies at FIR and radio ($3.5$ and $6.2$ cm)  show different configurations. The B-field traced at FIR wavelengths is located at the contact regions of the dust lane with the starburst ring along the East and West sides with a dominant power of $m=0$ at a level of $24$\%. This B-field also has the additional contributions of $m=1$ and $m=2$ at a $\sim20$\% level each. At radio wavelengths, the B-field has a dominant power of $m=2$ at a level of $30$\%, with the additional contributions of $m=1$ and $m=0$ at a $\sim20$\% level each. These results show that this technique, which is model-independent, can also be a powerful tool to disentangle the B-field morphologies of nearby galaxies and MHD simulations of galaxies.

We also performed a detailed study of the B-field morphology around the starburst ring and as a function of the distance from the central active nucleus using several tracers of the ISM (i.e. column density, dust temperature, and velocity dispersion of the molecular gas). All radial and azimuthal profiles were estimated in the plane of the sky. Results show that there is a spatial shift of several tracers across the starburst ring as a function of the distance to the AGN. The \lineco\ integrated emission line peaks at the inner $\sim7.5\arcsec$ ($0.69$ kpc) edge of the starburst ring, while the radio polarized flux peaks at the outer $\sim15\arcsec$ ($1.39$ kpc) edge. The FIR polarized flux peaks at a radius of $\sim11\arcsec$ ($1.02$ kpc) in the contact regions between the galactic bar and the starburst ring. We compared the FIR polarization with the kinematics of the molecular gas using the velocity dispersion of the \lineco\ emission line. We measured that a) the 89 \um\ polarization fraction is almost constant with velocity dispersion of the molecular gas, b) the molecular gas is not located with the star forming regions, and c) the peaks of 89 \um\ polarized flux do not coincide with the star forming regions. These result suggests that the FIR polarization is not directly correlated with the star formation activity in the starburst ring.  At radio wavelengths, we found no clear trends between the polarization or polarized flux with any of the aforementioned tracers.  The dust temperature is higher, $T_{89\,\mu \mathrm{m}} = 30.7\pm0.4$ K, at the position of the $89$ \um\ polarized flux than at the peaks at of the $3.5$ and $6.2$ cm polarized flux, $T_{3\,\mathrm{cm}} = 26.2\pm1.7$ K.

We propose a scenario where the dense ISM has a compressed B-field at the contact regions of the galactic bar and the starburst ring. The B-field is then deflected at the contact points due to the transition between the compressed B-field in the bar to a MHD dynamo towards the active nucleus. The B-field drags the dense gas towards the orbits of the starburst ring. The diffuse ionized ISM mixed with the relativistic electrons responds to shear flows, which drags the diffuse gas towards the active nucleus through a spiral-like structure via a galactic dynamo action. We have shown that both the FIR and radio polarimetric observations in combination with probes of the gas turbulence field, dust temperature, and gas density are critical ingredients to our understanding of the accretion flow from kpc-scale to hundreds pc-scale towards the active galactic nucleus of NGC 1097. Our results showing that the B-field inferred from radio polarimetric observations is similar to the gas kinematics in a bar potential, while the B-field inferred from FIR polarimetric observations resembles the flow in a shock are consistent with a MHD-driven flow within the central 1 kpc. We concluded that B-fields may be controlling the transfer of matter from the starburst ring formed in barred galaxies towards the active nuclei. Further multi-wavelength polarimetric observations of barred galaxies and MHD simulations are required to test whether our proposed scenario is general.

%What have we learned? What we were missing before? You say the gas flow follows the B-field morphology towards the active nucleus. Does this mean that the B-field is enabling/helping the gas to move (faster) towards the nucleus for example?


\begin{acknowledgments}

%We are grateful to the anonymous referee, whom comments greatly helped to clarify and improve the manuscript.  
Based on observations made with the NASA/DLR Stratospheric Observatory for Infrared Astronomy (SOFIA) under the 07\_0034 Program. SOFIA is jointly operated by the Universities Space Research Association, Inc. (USRA), under NASA contract NNA17BF53C, and the Deutsches SOFIA Institut (DSI) under DLR contract 50 OK 0901 to the University of Stuttgart. KT has received funding from the European Research Council (ERC) under the European Unions Horizon 2020 research and innovation programme under grant agreement No. 771282

\end{acknowledgments}


\vspace{5mm}
\facilities{SOFIA (HAWC+), ALMA, \textit{Herschel}, VLA.}


\software{\textsc{astropy} \citep{astropy}, 
\textsc{APLpy} \citep{aplpy},
\textsc{matplotlib} \citep{hunter2007},
          }

%\appendix

%\bibliography{references}



\end{document}